\documentclass[conference]{IEEEtran}
\IEEEoverridecommandlockouts
\usepackage{cite}
\usepackage{mathtools}
\usepackage{amsmath,amssymb,amsfonts}
\usepackage{algorithmic}
\usepackage{graphicx}
\usepackage{textcomp}
\usepackage{xcolor}
\usepackage{fancyhdr}
\usepackage{hyperref}
\def\BibTeX{{\rm B\kern-.05em{\sc i\kern-.025em b}\kern-.08em
    T\kern-.1667em\lower.7ex\hbox{E}\kern-.125emX}}
\begin{document}

\title{A survey of loss functions for semantic segmentation}

\author{\IEEEauthorblockN{Shruti Jadon}
\IEEEauthorblockA{\textit{IEEE Member}\\
shrutijadon@ieee.org
}
}

\maketitle
%copyright notice

\thispagestyle{plain}

  \fancypagestyle{plain}{

  \fancyhf{} % clear all header and footer fields

  \fancyfoot[L]{978-1-7281-9468-4/20/\$31.00~\copyright2020~IEEE} % change copyright notice here if required

  \renewcommand{\headrulewidth}{0pt}

  \renewcommand{\footrulewidth}{0pt}

  }
\begin{abstract}
Image Segmentation has been an active field of research as it has a wide range of applications, ranging from automated disease detection to self driving cars. In the past 5 years, various papers came up with different objective loss functions used in different cases such as biased data, sparse segmentation, etc. In this paper, we have summarized some of the well-known loss functions widely used for Image Segmentation and listed out the cases where their usage can help in fast and better convergence of a model. Furthermore, we have also introduced a new log-cosh dice loss function and compared its performance on NBFS skull-segmentation open source data-set with widely used loss functions. We also showcased that certain loss functions perform well across all data-sets and can be taken as a good baseline choice in unknown data distribution scenarios.
\end{abstract}


\section{Introduction}
Deep learning has revolutionized various industries ranging from software to manufacturing. Medical community has also benefited from deep learning. There have been multiple innovations in disease classification, example, tumor segmentation using U-Net and cancer detection using SegNet. Image segmentation is one of the crucial contribution of deep learning community to medical fields. Apart from telling that some disease exists it also showcases where exactly it exists. It has drastically helped in creating algorithms to detect tumors, lesions etc. in various types of medical scans.

Image Segmentation can be defined as classification task on pixel level. An image consists of various pixels, and these pixels grouped together define different elements in image. A method of classifying these pixels into the a elements is called semantic image segmentation. The choice of loss/objective function is extremely important while designing complex image segmentation based deep learning architectures  as they instigate the learning process of algorithm. Therefore, since 2012, researchers have experimented with various domain specific loss function to improve results for their datasets. In this paper we have summarized fifteen such segmentation based loss functions that have been proven to provide state of art results in different domains. These loss function can be categorized into 4 categories:
Distribution-based, Region-based, Boundary-based, and Compounded (Refer \ref{tab0}). We have also discussed the conditions to determine which objective/loss function might be useful in a scenario. Apart from this, we have proposed a new log-cosh dice loss function for semantic segmentation. To showcase its efficiency, we compared the performance of all loss functions on NBFS Skull-stripping dataset \cite{Puccio2016ThePC} and shared the outcomes in form of Dice Coefficient, Sensitivity, and Specificity. The code implementation is available at GitHub: \url{https://github.com/shruti-jadon/Semantic-Segmentation-Loss-Functions}.

\begin{figure}[htbp]
\centerline{\includegraphics[scale = 0.33]{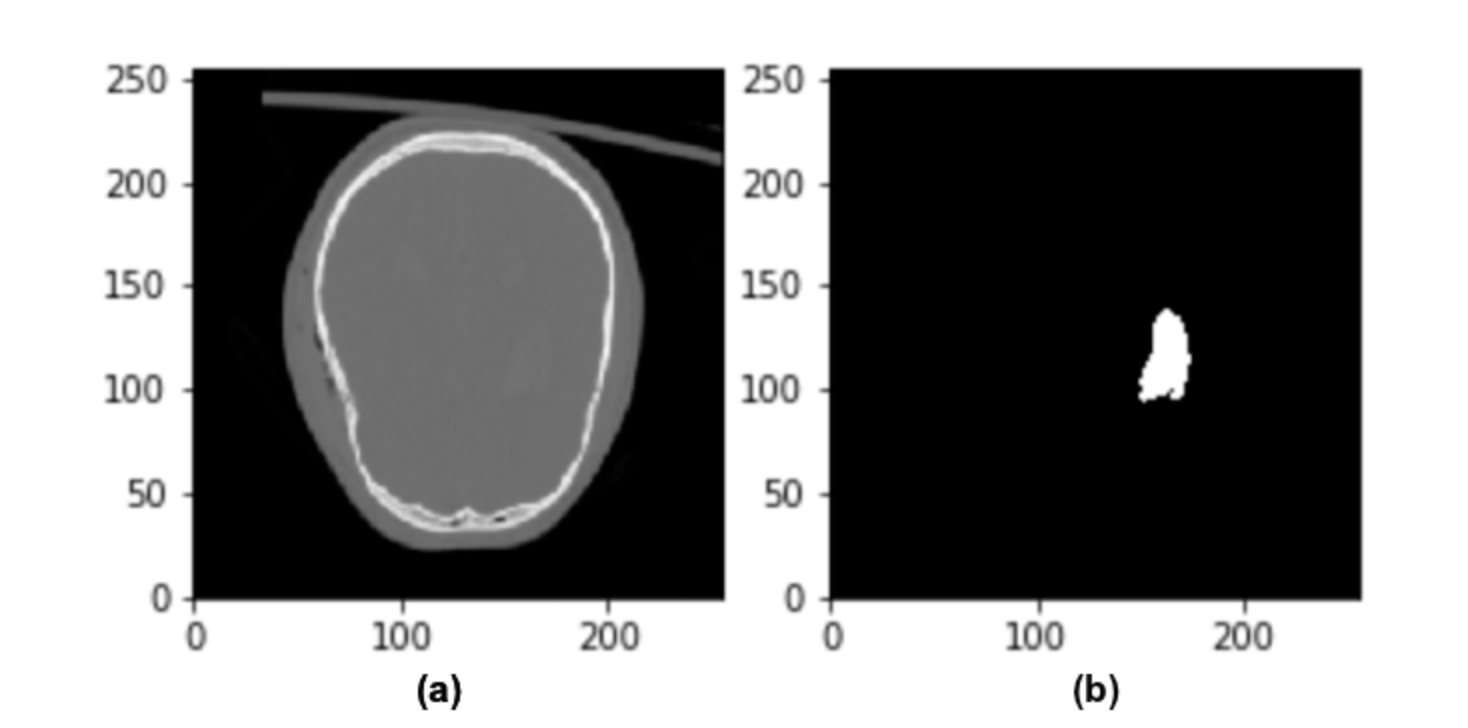}}
\caption{Sample Brain Lesion Segmentation CT Scan \cite{jadonspie20}. In this segmentation mask you can see, that number of pixels of white area(targeted lesion) is less than number of black pixels.}
\label{fig0}
\end{figure}

\begin{table}[htbp]
\caption{Types of Semantic Segmentation Loss Functions \cite{SegLossOdyssey}}
\begin{center}
\begin{tabular}{|c|c|}
\hline
\textbf{Type} & {\textbf{Loss Function}} \\
\hline
Distribution-based Loss & Binary Cross-Entropy\\
& Weighted Cross-Entropy\\
& Balanced Cross-Entropy\\
& Focal Loss\\
& Distance map derived loss penalty term \\
\hline
Region-based Loss & Dice Loss\\
& Sensitivity-Specificity Loss\\
& Tversky Loss\\
& Focal Tversky Loss\\
& \textbf{Log-Cosh Dice Loss}(ours)\\
\hline
Boundary-based Loss & Hausdorff Distance loss\\
& Shape aware loss\\
\hline
Compounded Loss & Combo Loss \\
& Exponential Logarithmic Loss \\
\hline
% \multicolumn{4}{l}{$^{\mathrm{a}}$Sample of a Table footnote.}
\end{tabular}
\label{tab0}
\end{center}
\end{table}

\section{Loss Functions}
Deep Learning algorithms use stochastic gradient descent approach to optimize and learn the objective. To learn an objective accurately and faster, we need to ensure that our mathematical representation of objectives, also known as loss functions are able to cover even the edge cases. The introduction of loss functions have roots in traditional machine learning, where these loss functions were derived on basis of distribution of labels. For example, Binary Cross Entropy is derived from Bernoulli distribution and Categorical Cross-Entropy from Multinoulli distribution. In this paper, we have focused on Semantic Segmentation instead of Instance Segmentation, therefore the number of classes at pixel level is restricted to 2. Here, we will go over 15 widely used loss functions and understand their use-case scenarios.

\begin{figure}[htbp]
\centerline{\includegraphics[scale = 0.2]{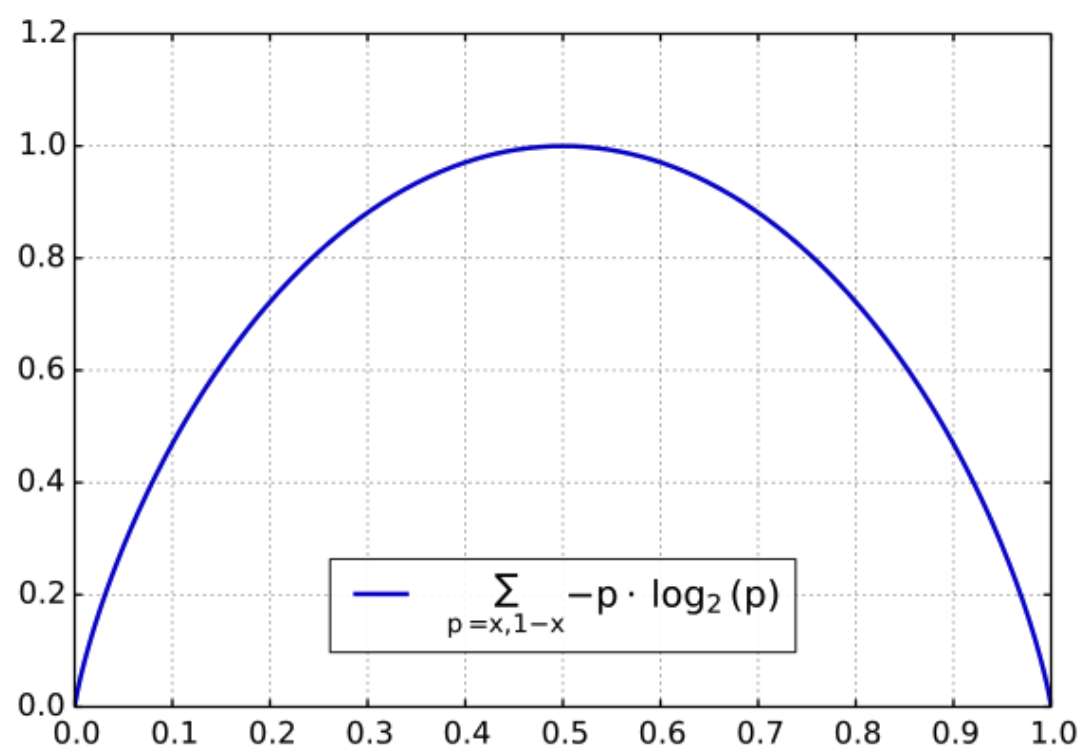}}
\caption{Graph of Binary Cross Entropy Loss Function. Here, Entropy is defined on Y-axis and Probability of event is on X-axis.}
\label{fig}
\end{figure}

\subsection{Binary Cross-Entropy}
Cross-entropy \cite{yi2004automated} is defined as a measure of the difference between two probability distributions for a given random variable or set of events. It is widely used for classification objective, and as segmentation is pixel level classification it works well.
\\
Binary Cross-Entropy is defined as:
\begin{equation}
    L_{BCE}(y,\hat{y})= -(ylog(\hat{y})+(1-y)log(1-\hat{y}))
\end{equation}
Here, $\hat{y}$ is the predicted value by the prediction model.

\subsection{Weighted Binary Cross-Entropy}
Weighted Binary cross entropy (WCE) \cite{pihur2007weighted} is a variant of binary cross entropy variant. In this the positive examples get weighted by some coefficient. It is widely used in case of skewed data \cite{ho2019real} as shown in figure 1. Weighted Cross Entropy can be defined as:
\begin{equation}
    L_{W-BCE}(y,\hat{y})= -(\beta*ylog(\hat{y})+(1-y)log(1-\hat{y}))
\end{equation}
\textbf{Note}: $\beta$ value can be used to tune false negatives and false positives. E.g; If you want to reduce the number of false negatives then set $\beta >1$, similarly to decrease the number of false positives, set $\beta<1$.

\subsection{Balanced Cross-Entropy}
Balanced cross entropy (BCE) \cite{xie2015holistically} is similar to Weighted Cross Entropy. The only difference is that in this apart from just positive examples \cite{pan2019diagnostic}, we also weight also the negative examples.
Balanced Cross-Entropy can be defined as follows: 
\begin{equation}
    L_{BCE}(y,\hat{y})= -(\beta*ylog(\hat{y})+(1-\beta)*(1-y)log(1-\hat{y}))    
\end{equation}
Here, $\beta$ is defined as $1 - \frac{y}{H*W}$

\subsection{Focal Loss}
Focal loss (FL) \cite{lin2002focal} can also be seen as variation of Binary Cross-Entropy. It down-weights the contribution of easy examples and enables the model to focus more on learning hard examples. It works well for highly imbalanced class scenarios, as shown in fig \ref{fig0}. Lets look at how this focal loss is designed. We will first look at binary cross entropy loss and learn how Focal loss is derived from cross-entropy. \\
\begin{equation}
  CE = \left \{
  \begin{aligned}
    &-log(p), && \text{if}\ y=1 \\
    &-log(1-p), && \text{otherwise}
  \end{aligned} \right.
\end{equation} 
To make convenient notation, Focal Loss defines the estimated probability of class as:
\begin{equation}
  p_t = \left \{
  \begin{aligned}
    & p, && \text{if}\ y=1 \\
    & 1-p, && \text{otherwise}
  \end{aligned} \right.
\end{equation} 
Therefore, Now Cross-Entropy can be written as,
\begin{equation}
   CE(p,y)=CE(p_t)=-log(p_t)
\end{equation}
Focal Loss proposes to down-weight easy examples and focus training on hard negatives using a modulating factor, $((1-p)t)^\gamma$ as shown below: 
\begin{equation}
    FL(p_t)=-\alpha_t (1-p_t)^\gamma log(p_t)
\end{equation}
Here, $\gamma>0$ and when $\gamma=1$ Focal Loss works like Cross-Entropy loss function. Similarly, $\alpha$ generally range from [0,1], It can be set by inverse class frequency or treated as a hyper-parameter. 

\subsection{Dice Loss}
The Dice coefficient is widely used metric in computer vision community to calculate the similarity between two images. Later in 2016, it has also been adapted as loss function known as Dice Loss \cite{sudre2017generalised}.
\begin{equation}
    DL(y,\hat{p})=1-\frac{2y\hat{p}+1}{y+\hat{p}+1}
\end{equation}
Here, 1 is added in numerator and denominator to ensure that the function is not undefined in edge case scenarios such as when $y=\hat{p}=0$.

\subsection{Tversky Loss}
Tversky index (TI) \cite{salehi2017tversky} can also be seen as an generalization of Dice’s coefficient. It adds a weight to FP (false positives) and FN (false negatives) with the help of $\beta$ coefficient.
\begin{equation}
    TI(p,\hat{p})=\frac{p\hat{p}}{p\hat{p}+\beta(1-p)\hat{p}+(1-\beta)p(1-\hat{p})}    
\end{equation}
Here, when $\beta=1/2$, It can be solved into regular Dice coefficient. Similar to Dice Loss, Tversky loss can also be defined as: 

\begin{equation}
    TL(p,\hat{p})=1- \frac{1+p\hat{p}}{1+p\hat{p}+\beta(1-p)\hat{p}+(1-\beta)p(1-\hat{p})}    
\end{equation}

\subsection{Focal Tversky Loss}
Similar to Focal Loss, which focuses on hard example by down-weighting easy/common ones. Focal Tversky loss \cite{abraham2019novel} also attempts to learn hard-examples such as with small ROIs(region of interest) with the help of $\gamma$ coefficient as shown below: 

\begin{equation}
    FTL= \sum_{c}(1-TI_c)^\gamma    
\end{equation}
here, $TI$ indicates tversky index, and $\gamma$ can range from [1,3].

\subsection{Sensitivity Specificity Loss}
Similar to Dice Coefficient, Sensitivity and Specificity are widely used metrics to evaluate the segmentation predictions. In this loss function, we can tackle class imbalance problem using $w$ parameter. The loss \cite{hashemi2018asymmetric} is defined as: 
\begin{equation}
    SSL=w*sensitivity+(1-w)*specificity    
\end{equation}
where,
\begin{equation}
    sensitivity=\frac{TP}{TP+FN}    
\end{equation}
and
\begin{equation}
    specificity=\frac{TN}{TN+FP}
\end{equation}

\subsection{Shape-aware Loss}
Shape-aware loss \cite{hayder2016shape} as the name suggests takes shape into account. Generally, all loss functions work at pixel level, however, Shape-aware loss calculates the average point to curve Euclidean distance among points around curve of predicted segmentation to the ground truth and use it as coefficient to cross-entropy loss function. It is defined as follows: 
\begin{equation}
    E_i=D(\hat{C},C_{GT})    
\end{equation}
\begin{equation}
    L_{shape-aware}=-\sum_i CE(y,\hat{y})-\sum{i}E_i CE(y,\hat{y})    
\end{equation}
Using $E_i$ the network learns to produce a prediction masks similar to the training shapes.

\subsection{Combo Loss}
Combo loss \cite{taghanaki2019combo} is defined as a weighted sum of Dice loss and a modified cross entropy. It attempts to leverage the flexibility of Dice loss of class imbalance and at same time use cross-entropy for curve smoothing. It's defined as: 

\begin{equation}
    L_{m-bce}=-\frac{1}{N}\sum_{i}\beta(y-log(\hat{y}))+(1-\beta)(1-y)log(1-\hat{y})
\end{equation}
\begin{equation}
    CL(y,\hat{y})=\alpha L_{m-bce} -(1-\alpha)DL(y,\hat{y})
\end{equation}
Here DL is Dice Loss.

\subsection{Exponential Logarithmic Loss}
Exponential Logarithmic loss \cite{wong20183d} function focuses on less accurately predicted structures using combined formulation of Dice Loss and Cross Entropy loss. Wong et al. \cite{wong20183d} proposes to make exponential and logarithmic transforms to both Dice loss an cross entropy loss so as to incorporate benefits of finer decision boundaries and accurate data distribution. It is defined as: 

\begin{equation}
    L_{Exp}=w_{Dice}L_{Dice}+w_{cross}L_{cross}
\end{equation}
where
\begin{equation}
    L_{Dice}=E(-ln(DC)^{\gamma_{Dice}})
\end{equation}

\begin{equation}
    L_{cross}=E(w_l(-ln(p_l))^{\gamma_{cross}}))
\end{equation}
 Wong et al. \cite{wong20183d} have used $\gamma_{cross}=\gamma_{Dice}$ for simplicity.

\subsection{Distance map derived loss penalty term}
Distance Maps can be defined as distance (euclidean, absolute, etc.) between the ground truth and the predicted map. There are two ways to incorporate distance maps, either create neural network architecture where there's a reconstruction head along with segmentation, or induce it into loss function. Following same theory, Caliva et al. \cite{caliva2019distance} have used distance maps derived from ground truth masks and created a custom penalty based loss function. Using this approach, its easy to guide the network’s focus towards hard-to-segment boundary regions. The loss function is defined as: 

\begin{equation}
    L(y,p)=\frac{1}{N}\sum_{i=1}^{N}(1+\phi) (\odot) L_{CE}(y,p)
\end{equation}
Here, $\phi$ are generated distance maps 

\textbf{Note} Here, constant $1$ is added to avoid vanishing gradient problem in U-Net and V-Net architectures.

\begin{figure}[htbp]
\centerline{\includegraphics[scale = 0.5]{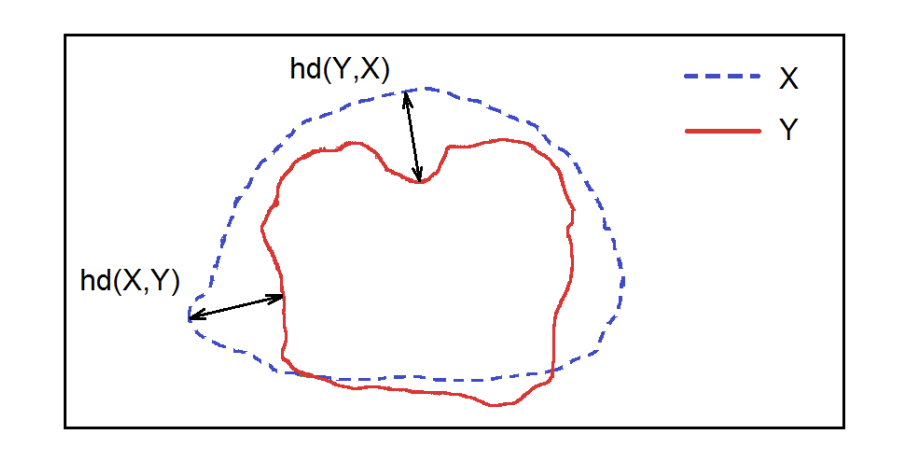}}
\caption{Hausdorff Distance between point sets X and Y \cite{karimi2019reducing}}
\label{fig31}
\end{figure}

\subsection{Hausdorff Distance Loss}
Hausdorff Distance (HD) is a metric used by segmentation approaches to track the performance of a model. It is defined as: 

\begin{equation}
    d(X,Y)=max_{x \epsilon X}min_{y \epsilon Y} ||x-y||_2    
\end{equation}
The objective of any segmentation model is to maximize the Hausdorff Distance \cite{Ribera2018WeightedHD}, but due to its non-convex nature, its not widely used as loss function. Karimi et al. \cite{karimi2019reducing} has proposed 3 variants of Hausdorff Distance based loss functions which incorporates the metric use case and ensures that the loss function is tractable. These 3 variants are designed on basis of how we can use Hausdorff Distance as part of loss function: (i) taking max of all HD errors, (ii) minimum of all errors obtained by placing a circular structure of radius r, and (iii) max of a convolutional kernel placed on top of missing segmented pixels.

\subsection{Correlation Maximized Structural Similarity Loss}
A lot of semantic based segmentation loss functions focus on classification error at pixel level while disregarding the pixel level structural information. Some other loss functions \cite{zhao2019correlation} have attempted to add information using structural priors such as CRF, GANs, etc. In this loss functions, zhao et al. \cite{zhao2019correlation} have introduced a Structural Similarity Loss (SSL) to achieve a high positive linear correlation between the
ground truth map and the predicted map. Its divided into 3 steps: Structure Comparison, Cross-Entropy weight coefficient determination, and mini-batch loss definition. \\
As part of Structure comparison, authors have calculated e-coefficient, which can measure the degree of linear correlation between ground truth and prediction: 

\begin{equation}
    e = |\frac{y-\mu_y+C_4}{\sigma_y+C_4} - \frac{p-\mu_p+C_4}{\sigma_p+C_4}|
\end{equation}

Here, $C_4$ is stability factor set to 0.01 as an empirical observed value.
$\mu_y$ and $\sigma_y$ are the local mean and standard deviation
of the ground truth y respectively. y locates at the center of the local region and p is the predicted probability. \\
After calculating the degree of correlation, zhao et al. \cite{zhao2019correlation} have used it as coefficient for cross entropy loss function, defined as: \\
\begin{equation}
    f_{n,c}=1*{e_{n,c}>\beta e_{max}}
\end{equation}

Using this coefficient function, we can define SSL loss as:

\begin{equation}
    Loss_{ssl}(y_{n,c},p_{n,c})=e_{n,c}f_{n,c}L_{CE}(y_{n,c},p_{n,c})    
\end{equation}

and finally for mini-batch loss calculation, The SSL can be defined as: 
\begin{equation}
    L_{ssl}=\frac{1}{M}\sum_{n=1}^{N}\sum_{c=1}^{C}L_{ssl}(y_{n,c},p_{n,c})    
\end{equation}
where, M is $\sum_{n=1}^{N}\sum_{c=1}^{C}f_{n,c}$
Using above formula, loss function will automatically abandon those pixel level predictions, which doesn't show correlation in terms of structure.
\begin{table*}[htbp]
\caption{Tabular Summary of Semantic Segmentation Loss Functions}
\begin{center}
\begin{tabular}{|c|c|}
\hline
\textbf{Loss Function} & {\textbf{Use cases}} \\
\hline
Binary Cross-Entropy & Works best in equal data distribution among classes scenarios\\
& Bernoulli distribution based loss function \\
\hline
Weighted Cross-Entropy & Widely used with skewed dataset\\
& Weighs positive examples by $\beta$ coefficient\\
\hline
Balanced Cross-Entropy & Similar to weighted-cross entropy, used widely with skewed dataset\\
& weighs both positive as well as negative examples by $\beta$ and $1-\beta$ respectively\\
\hline
Focal Loss & works best with highly-imbalanced dataset\\
& down-weight the contribution of easy examples, enabling model to learn hard examples\\
\hline
Distance map derived loss penalty term & Variant of Cross-Entropy\\
& Used for hard-to-segment boundaries\\
\hline
Dice Loss & Inspired from Dice Coefficient, a metric to evaluate segmentation results. \\
& As Dice Coefficient is non-convex in nature, it has been modified to make it more tractable.\\
\hline
 Sensitivity-Specificity Loss & Inspired from Sensitivity and Specificity metrics\\
 & Used for cases where there is more focus on True Positives.\\
\hline
 Tversky Loss & Variant of Dice Coefficient\\
 & Add weight to False positives and False negatives.\\
\hline
 Focal Tversky Loss & Variant of Tversky loss with focus on hard examples\\
 & \\
\hline
Log-Cosh Dice Loss(ours) & Variant of Dice Loss and inspired regression log-cosh approach for smoothing\\
 & Variations can be used for skewed dataset\\
\hline
Hausdorff Distance loss & Inspired by Hausdorff Distance metric used for evaluation of segmentation\\
& Loss tackle the non-convex nature of Distance metric by adding some variations\\
\hline
Shape aware loss & Variation of cross-entropy loss by adding a shape based coefficient\\
& used in cases of hard-to-segment boundaries.\\
\hline
Combo Loss & Combination of Dice Loss and Binary Cross-Entropy\\
& used for lightly class imbalanced by leveraging benefits of BCE and Dice Loss\\
\hline
Exponential Logarithmic Loss & Combined function of Dice Loss and Binary Cross-Entropy\\
& Focuses on less accurately predicted cases\\
\hline
Correlation Maximized Structural Similarity Loss & Focuses on Segmentation Structure. \\
& Used in cases of structural importance such as medical images.\\
\hline
% \multicolumn{4}{l}{$^{\mathrm{a}}$Sample of a Table footnote.}
\end{tabular}
\label{tab2}
\end{center}
\end{table*}

\subsection{Log-Cosh Dice Loss}
Dice Coefficient is a widely used metric to evaluate the segmentation output. It has also been modified to be used as loss function as it fulfills the mathematical representation of segmentation objective. But due to its non-convex nature, it might fail in achieving the optimal results. Lovász-Softmax loss \cite{berman2017lovszsoftmax} aimed to tackle the problem of non-convex loss function by adding the smoothing using Lovász extension. Log-Cosh approach has been widely used in regression based problem for smoothing the curve.

\begin{figure}[htbp]
\centerline{\includegraphics[scale = 0.15]{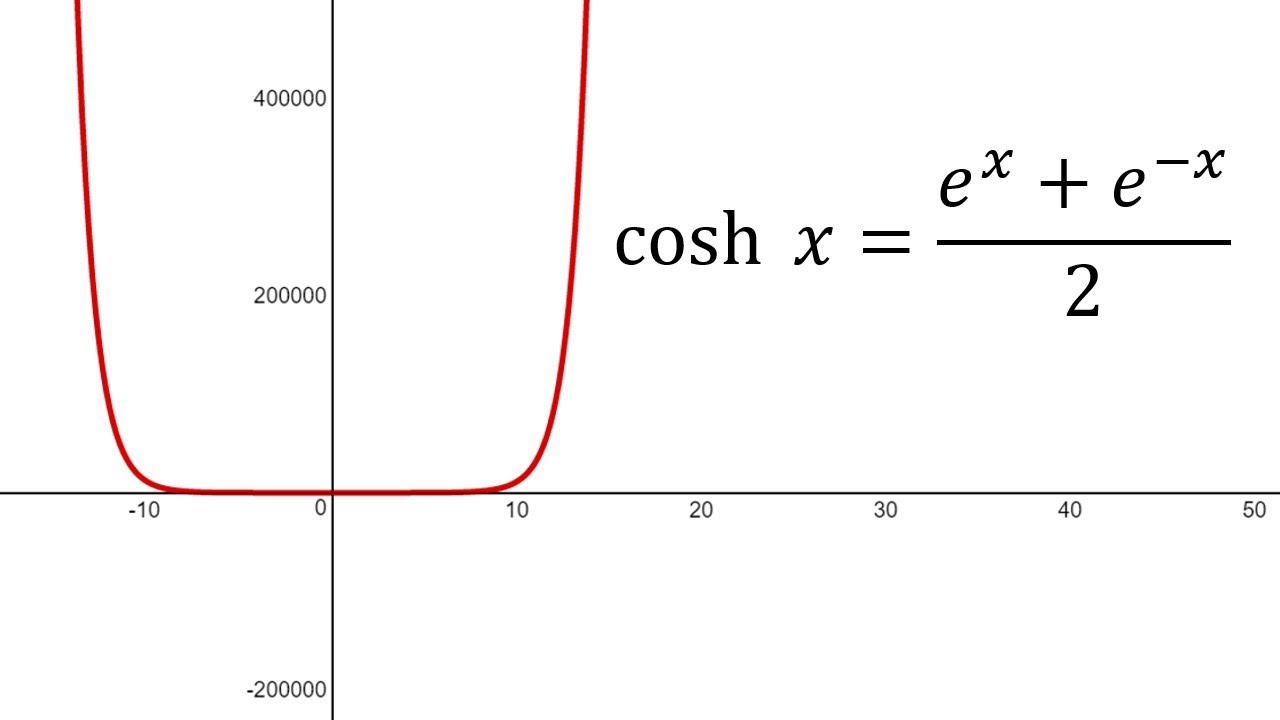}}
\caption{Cosh(x) function is the average of $e^x$ and $e^-x$ }
\label{fig1}
\end{figure}

\begin{figure}[htbp]
\centerline{\includegraphics[scale = 0.3]{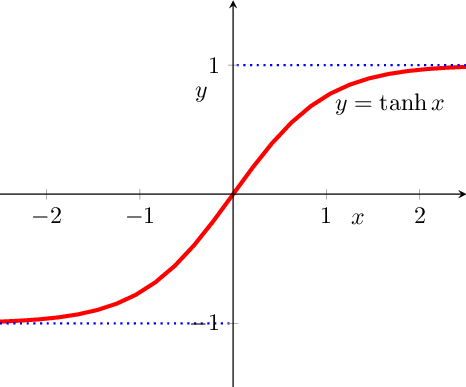}}
\caption{tanh(x) function is continuous and finite. It ranges from $[-1,1]$}
\label{fig3}
\end{figure}

Hyperbolic functions have been used by deep learning community in terms of non-linearities such as tanh layer. They are tractable as well as easily differentiable. Cosh(x) is defined as (ref \ref{fig1}):

\begin{equation}
    cosh x=\frac{e^x+e^{-x}}{2}
\end{equation}
and
\begin{equation}
    cosh'x=\frac{e^x-e^{-x}}{2}=sinh x
\end{equation}
but, at present $cosh x$ range can go up to infinity. So, to capture it in range, $log$ space is used, making the log-cosh function to be:
\begin{equation}
    L(x)=log(cosh x)   
\end{equation}
and using chain rule 
\begin{equation}
    L'(x)=\frac{sinh x}{cosh x}=tanh x
\end{equation}
which is continuous and finite in nature, as $tanh x$ ranges from $[-1,1]$

\begin{figure}[htbp]
\centerline{\includegraphics[scale = 0.7]{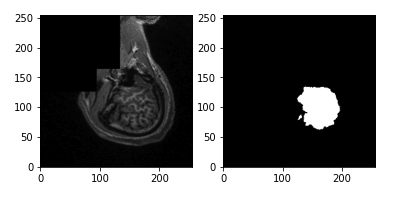}}
\caption{Sample CT scan image from NBFS Skull Stripping Dataset \cite{Puccio2016ThePC}}
\label{fig2}
\end{figure}

\begin{table}[htbp]
\caption{Comparison of some above mentioned loss functions on basis of Dice scores, Sensitivity and Specificity for Skull Segmentation}
\begin{center}
\begin{tabular}{|c|c|c|c|}
\hline
\textbf{Loss}&\multicolumn{3}{|c|}{\textbf{Evaluation Metrics}} \\
\cline{2-4} 
\textbf{Functions} & \textbf{\textit{Dice Coefficient}}& \textbf{\textit{Sensitivity}}& \textbf{\textit{Specificity}} \\
\hline
Binary Cross-Entropy & 0.968 & 0.976 & 0.998 \\
Weighted Cross-Entropy & 0.962 & 0.966 & 0.998 \\
Focal Loss &0.936 & 0.952&\textbf{0.999}  \\
Dice Loss & 0.970 & 0.981 & 0.998  \\
Tversky Loss & 0.965 & 0.979 & 0.996 \\
\textbf{Focal Tversky Loss} & \textbf{0.977} &\textbf{0.990} &0.997  \\
Sensitivity-Specificity Loss & 0.957 &0.980 &0.996  \\
Exp-Logarithmic Loss & 0.972 &0.982 &0.997  \\
\textbf{Log Cosh Dice Loss} & \textbf{0.989} & 0.975 & 0.997  \\
\hline
% \multicolumn{4}{l}{$^{\mathrm{a}}$Sample of a Table footnote.}
\end{tabular}
\label{tab1}
\end{center}
\end{table}

On basis of above proof which showcased that Log of Cosh function will remain continuous and finite after first order differentiation. We are proposing Log-Cosh Dice Loss function for its tractable nature while encapsulating the features of dice coefficient. It can defined as: 
\begin{equation}
    L_{lc-dce}=log(cosh(Dice Loss))
\end{equation}

\section{Experiments}
For experiments, we have implemented simple 2D U-Net model \cite{jadonspie20} architecture for segmentation with 10 convolution encoded layers and 8 decoded convolutional transpose layers. We have used NBFS Skull-stripping dataset\cite{Puccio2016ThePC}, which consists of ~125 skull CT scans, and each scan consists of ~120 slices (refer figure \ref{fig2}). For training, we have used batch size of 32 and adam optimizer with learning rate 0.001 and learning rate reduction up to $10^{-8}$. As part of training, validation, and test data, we have split data-set into 60-20-20. We have performed experiments using only 9 loss functions as other loss functions were either resolving into our existing chosen loss function or weren't fit for NBFS skull dataset. After training the model for different loss functions, we have evaluated them on basis of well known evaluation metrics: Dice Coefficient, Sensitivity, and Specificity.
\subsubsection{Evaluation Metrics}

Evaluation Metrics plays an important role in assessing the outcomes of segmentation models. In this work, we have analyzed our results using Dice Coefficient, Sensitivity, and Specificity metric. Dice Coefficient, also known as overlapping index measures the overlapping between ground truth and predicted output. Similarly, Sensitivity gives more weightage to True Positives and Sensitivity calculates the ratio of True Negatives. Collectively, these metrics examine the model performance effectively.
\begin{equation}
DC = \frac{2TP}{2TP+FP+FN},
\end{equation}
\begin{equation}
Sensitivity(TPR) = \frac{TP}{TP+FN}, and
\end{equation}
\begin{equation}
Specificity(TNR) = \frac{TN}{TN+FP}
\end{equation}
In Conclusion, by using ~40,000 annotated segmented examples, we achieved an optimal dice coefficient of ~0.98 using Focal Tversky Loss. Log-Cosh Dice Loss function also achieved similar results of the dice coefficient ~0.975, very close to the best results. As of sensitivity, i.e., True Positive Rate, Focal Tversky Loss outperformed all other loss functions, whereas specificity(True Negative Rate) remained consistent across all loss functions. We have also observed similar outcomes in our past research \cite{jadonspie20} Focal Tversky loss and Tversky loss generally gives optimal results with right parameter values.
\section{Conclusion}
Loss functions play an essential role in determining the model performance. For complex objectives such as segmentation, it's not possible to decide on a universal loss function. The majority of the time, it depends on the data-set properties used for training, such as distribution, skewness, boundaries, etc. None of the mentioned loss functions have the best performance in all the use cases. However, we can say that highly imbalanced segmentation works better with focus based loss functions.
Similarly, binary-cross entropy works best with balanced data-sets, whereas mildly skewed data-sets can work around smoothed or generalized dice coefficient. In this paper, we have summarized 14 well-known loss functions for semantic segmentation and proposed a tractable variant of dice loss function for better and accurate optimization. In the future, we will use this work as a baseline implementation for few-shot segmentation \cite{jadon2019hands} experiments.

\small
 \nocite{*} % to test all bib entrys
\bibliographystyle{unsrt}
\bibliography{egbib}

\end{document}